# Imaging Local Sources of Intermodulation in Superconducting Microwave Devices

A. P. Zhuravel, A. V. Ustinov, D. Abraimov, and Steven M. Anlage, *Member, IEEE*

*Abstract*—This work presents new experimental results on low-temperature (LT) characterization of local rf properties of passive superconducting (SC) microwave devices using a novel Laser Scanning Microscope (LSM). In this technique, a modulated laser beam is focused onto and scanned over the surface of a resonant SC device to probe the spatial distribution of rf current. The highly localized photo-induced change of the kinetic inductance of the SC device produces both a shift of the resonant frequency $f_0$ and change of the quality factor Q. An image of these changes is recorded as the laser spot is scanned over the device. We present the first measurements of spatially resolved intermodulation response in a High Temperature Superconducting (HTS) co-planar waveguide resonator, opening up a new window into the local origins of nonlinearity in the HTS materials.

## I. Introduction

SUPERCONDUCTING microwave devices are currently of great interest due to their extensive applications in modern communication technologies. Passive microwave circuits consisting of high-$T_c$ superconductors (HTS) are being used as delay lines, multiplexers, resonators and filters for mobile, cellular and satellite communications [1]-[3]. One of the crucial problems in applying HTS materials remains their relatively large and inhomogeneous nonlinear surface impedance [4]-[5]. This behavior is due, in part, to inhomogeneous current flow and leads to nonlinearity of the device response with respect to applied power.

The *microscopic* mechanisms for nonlinearity have not yet been definitively identified. The conventional methods of analyzing the nonlinear response are based on *global* characteristics of the sample, such as intermodulation distortion (IMD), generation of harmonics, and nonlinear surface impedance [6]-[11]. A macroscopic structure/property study concluded that "lattice distortions of the a-b plane in grains and grain boundaries" is responsible for enhanced IMD in YBCO films [4]. This kind of analysis does not directly lead to a solution of the problem, i.e. to finding the origin of the observed nonlinearity.

Various kinds of microscopic techniques have been employed to identify extrinsic sources of nonlinearity [12]-[14]. Among these efforts, the LSM observations [12], [13] suffered from modest spatial resolution and failed to see an expected re-distribution of current in an HTS resonator as the rf power increased [12]. Room temperature modulated optical reflectance measures the local carrier density, and can be used to find defects that affect the superconducting microwave performance. [14] A previous attempt to measure the local origins of nonlinearity was made by Hu, *et al*. [15] They used a superconducting microstrip resonator and measured the rf electric field above the device with a scanned coaxial probe. Their images of the IMD signal where a global superposition of signals generated throughout the device, and did not point to the origins of the signals.

It has been shown that IMD measurements are much more sensitive to nonlinearities than are Q and frequency shift measurements, and that these signals are non-thermal in origin [16]. In this paper we present a new technique for possible microscopic identification of sources of IMD in superconducting microwave devices, based on LSM imaging of operating devices.

## II. Experimental Details

Measurements were performed on an HTS co-planar waveguide (CPW) resonator that was fabricated from a 240 nm thick YBa$_2$Cu$_3$O$_{7-\delta}$ (YBCO) film deposited on a 500 μm thick LaAlO$_3$ (LAO) substrate by laser ablation. The CPW has a strip line of 500 μm width and 7.75 mm length. The line was separated from the ground planes by 650 μm and coupled to the feed lines via two capacitive gaps 500 μm wide. The 20x10x0.5 mm sample chip was glued by vacuum grease to a brass microwave package. Silver paste was applied to the Au contacts laser ablated onto the ground planes. Electrical contacts were made by spring loading the center pins of the SMA connectors to the corresponding Au contacts on the center conductor lines through a thin indium layer. The microwave package was screwed to a copper cold stage inside the vacuum cavity of a variable-temperature optical cryostat. This cryostat stabilized the temperature of the sample in the

Manuscript received August 5, 2002. This work was supported in part by a NATO Collaborative Linkage Grant No. 431679 from the NATO Scientific and Environmental Affairs Office, the Maryland Center for Superconductivity Research, Maryland/Rutgers/NSF MRSEC DMR-00-80008, and NSF/Neocera SBIR-II DMI-00-78486.

Alexander P. Zhuravel is with B. Verkin Institute for Low Temperature Physics & Engineering, National Academy of Sciences of Ukraine, 61164 Kharkov, Ukraine (phone: +380-572-308507; fax: +380-572-322370; e-mail: zhuravel@ilt.kharkov.ua).

Alexey V. Ustinov and Dmytro V. Abraimov are with Physics Institute III, University of Erlangen-Nuremberg, D-91058, Erlangen, Germany (e-mail: ustinov@physik.uni-erlangen.de).

Steven M. Anlage is with the Physics Department, Center for Superconductivity Research, University of Maryland, College Park, MD 20742-4111 USA (e-mail: anlage@squid.umd.edu).

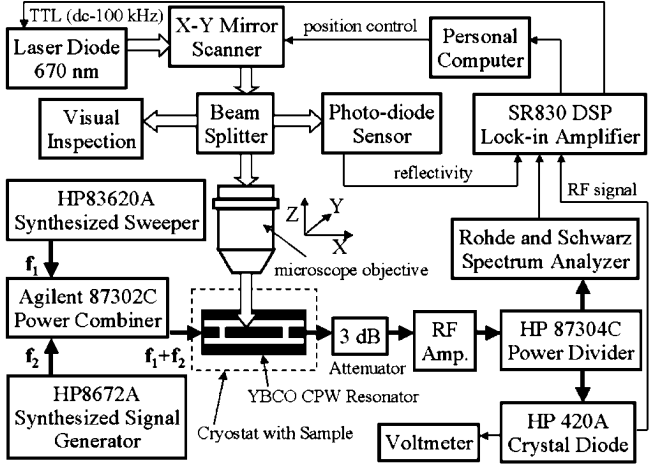

Fig. 1. Schematic diagram of IMD LSM microscope optics and the microwave electronics used for IMD imaging. Open arrows show the optical beam path while dark filled arrows show the microwave signal.

range *6-300 K* with an accuracy of *5 mK* by using a resistive heater on the cold stage. The transition temperature of the YBCO film was measured to be *86 K*. The fundamental resonant frequency $f_0$ for the resonator at 78 K was approximately $f_0$ = 5.2 GHz with a loaded $Q$ ~ 2500.

A schematic of the optics and microwave electronics of our IMD LSM setup is illustrated in Figure 1. The general principle of operation of the LSM is to scan the surface of an SC film with a tightly-focused laser beam (probe) for 2-D reconstruction of the response signal $\delta V(x,y)$ arising from local laser-sample interaction. Any changes, either in laser beam optical characteristics or in electronic transport of the excited SC sample due to heating and/or direct depairing of the Cooper pair condensate underneath the probe, contributes to the response signal.

A 20x long-working-distance (20 mm) microscope objective with a numerical aperture (NA) of 0.42 is used to focus the 5mW light probe on the sample from a laser diode emitting a 670 nm wavelength beam. It is raster scanned (over a 250 μm x 250 μm area) by two closely spaced orthogonally oriented mirror galvanometers by equal steps ranging from 0.1 to 2 μm on the sample surface. The beam is Gaussian shaped by a single-mode optical fiber and amplitude modulated up to 100 kHz by a TTL signal from the lock-in oscillator.

Three LSM operating modes were employed, depending on the photo-response (PR) mechanism exploited for imaging. First, the reflectance-mapping mode was applied to image the sample topology, as well as the position, shape and size of visible defects (irregularities) in both the substrate and the HTS film. In this mode, the modulated laser beam was reflected from the sample surface and monitored by an optical sensor to produce an ac electrical signal proportional to the local sample reflectivity as a function of probe position.

The second (rf current imaging) mode produces a map of local rf current density squared. A synthesized signal generator (transmitted microwave power = ~ -15 to 0 dBm) excited a resonant mode of the CPW resonator. The local heating of the sample by the "hot-spot" of a focused laser beam shifted both $f_0$ and $Q$ of the device due to changes in the local magnetic penetration depth and the stored energy in the resonator [12, 13, 17]. This caused a change in the $S_{12}(f)$ transmission curve that is proportional to the local rf current density squared $[J_{RF}^2(x,y)]$ and leads to the change in transmitted power $P$ given by [12,18]:

$$\delta P \sim (\lambda J_{RF}(x,y))^2 A \delta\lambda , \quad (1)$$

where $\lambda$ is the magnetic penetration depth, $J_{RF}(x,y)$ is the current density at the site of the perturbation, $A$ is the "spot size" and $\delta\lambda$ is the change in penetration depth caused by modulated laser heating. A crystal diode detects the rf amplified changes in laser-modulated rf power, and an image of these changes is recorded as a function of the location of the laser spot [19].

In the third (IMD imaging) mode, two fixed frequency signals ($f_1$ and $f_2$) were applied to the CPW resonator. The $f_1$ and $f_2$ were centered on the $|S_{12}(f)|$ curve with a spacing of 10 kHz and had the same amplitudes. Changes in $P_{f1}$ or $P_{2f1-f2}$ as a function of position $(x,y)$ of the laser beam perturbation on the sample were imaged. A spectrum analyzer was used to measure the power in the tones ($P_f$) at these intermodulation frequencies to see how the global nonlinear response was changed by the local perturbation. The change in IMD transmitted power $P_{2f1-f2}$ is estimated to be:

$$\frac{\delta P_{2f_1-f_2}}{P_{2f_1-f_2}} \sim \left\{ \frac{\delta\lambda}{\lambda} - \frac{\delta J_{IMD}}{J_{IMD}} - \frac{\int \delta\lambda\, R_s\, J_{RF}^2\, dS + \int \delta R_s\, \lambda\, J_{RF}^2\, dS}{\int \lambda\, R_s\, J_{RF}^2\, dS} \right\} \quad (2)$$

where $R_s$ is the surface resistance, $J_{IMD}$ is the nonlinearity current scale,[8, 11] and $\delta J_{IMD}$ is the change in nonlinearity current scale caused by the laser heating. Note that this treatment of nonlinearity (through $J_{IMD}$) is general and applies equally well to intrinsic and extrinsic sources. Hence the LSM IMD photoresponse is related to changes in the local nonlinearity current scale as well as changes in penetration depth and surface resistance at the site of the perturbation.

We measure the change in global IMD produced by heating a small area of the sample. We shall assume that the more nonlinear parts of the material will contribute a bigger change to the IMD power when they are heated. Based on numerical simulations of Eq. (2) with the two-fluid model, we find to first approximation that the contrast seen by the LSM tuned to the intermodulation frequency is proportional to the local change in intermodulation current density scale, $J_{IMD}$.

### III. RESULTS AND DISCUSSION

#### A. Lateral resolution

Optical as well as thermal aberrations both limit the spatial resolution $S$ of the LTLSM technique when probing electronic properties of superconductors. Roughly, the smallest features of scanned LSM images may be resolved on a length-scale of $S=(d_{opt}^2+l_T^2)^{1/2}$, where $d_{opt}$ is the optical resolution [in terms of the full-width at half-maximum (FWHM) of the focused Gaussian laser beam], and $l_T$ is the thermal healing length.

As a test to determine $d_{opt}$, the sharp edge of a Au pad on LAO substrate was imaged. Figure 2 shows the profile (open circles) of local reflectance variation measured by the photo-diode sensor along a 20 μm line scan through the Au/LAO interface with 0.1 μm steps of the scanning laser probe. The Gaussian fit (solid line) of the edge-scan derivative (open squares) gives $\sigma = 0.585 +/- 0.003$ μm, i.e. $d_{opt} \sim$ FWHM = $2\sigma(2\ln 2)^{1/2} = 1.38$ μm for the laser spot.

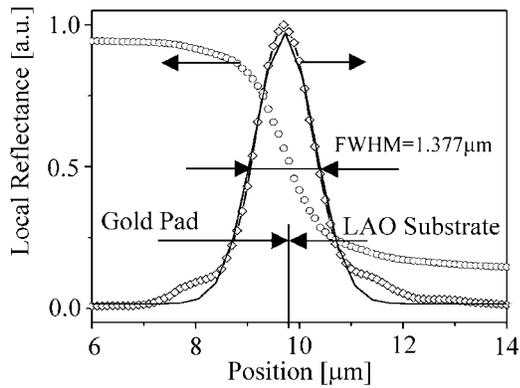

Fig. 2. Measured line-scan profile of the reflection from the edge of a Au contact pad on LAO substrate and its derivative fit (solid line) to a Gaussian; scanning step 100 nm

To estimate $l_T$ we analyzed the thermal diffusion of heat away from the laser probe during one cycle of the modulation frequency $f_m$ of incident laser power. Due to good acoustic mismatch between the HTS film and substrate at *78 K*, the thermal healing length can be expressed as [20]: $l_T = (k/c\rho_m f_m)^{1/2}$, where $\rho_m$ is the mass density, $c$ is the specific heat, and $k$ is the thermal conductivity of both LAO and YBCO acting in parallel. By probing the modulation-frequency-dependent distribution of the kinetic-inductance PR at the edge of the superconducting strip we directly measured the value of $S = 4.2$ μm at $f_m = 100$ kHz to $S = 40.1$ μm at 1 kHz. Hence, assuming that no other parameters vary significantly with probe position, our estimate $l_T (f_m)$ value varies from 4 μm to 40 μm at these $f_m$, in good agreement with values in the literature [21].

*B. Intermodulation Imaging*

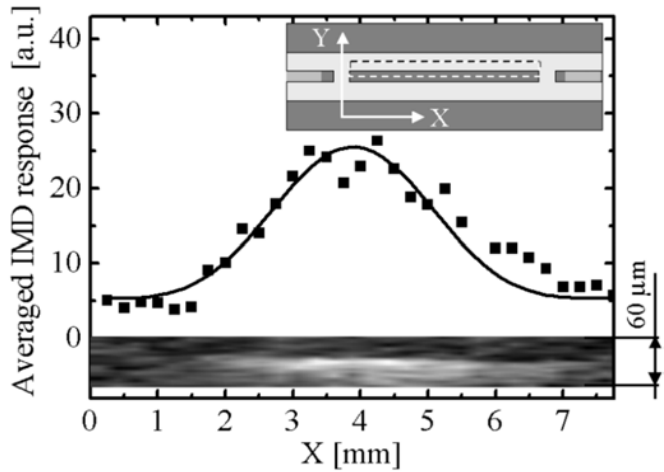

Fig. 3. LTLSM 7.75 mm x 60 μm image showing the distribution of resonator IMD PR of the CPW center strip along one edge marked by the box in inset. White and black regions correspond to sample areas with the maximum and zero IMD signal, respectively. The solid line through measured data (solid squares) is fit to $\cos^6(kx+\phi)$. The data points were obtained by averaging RF PR in transverse line-cuts in 250 μm steps along the center strip.

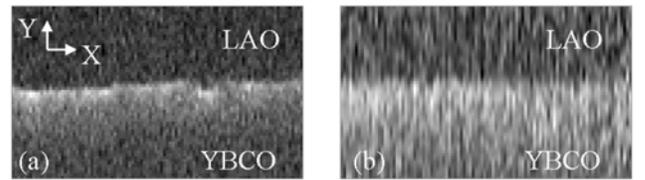

Fig. 4. 2-D LTLSM image of (a) RF current PR and (b) IMD photo-response distribution in a 100 μm x 60 μm area along one edge near the center of the device. Brighter regions correspond to larger response

Images of the ordinary PR (imaging mode 2 above) of the entire CPW resonator show a clear $\cos^2(kx+\phi)$ dependence along the length of the resonator, as expected [12,19]. Here we present the first preliminary images of IMD PR (imaging mode 3) on the same resonator. Figure 3 shows IMD PR along one edge and along the entire length of the CPW center strip. Note that the IMD PR is much more inhomogeneous than the ordinary rf PR [19]. A longitudinal line cut through the data shows that the IMD PR can be fit to the expected $\cos^6(kx+\phi)$ form. The data shows strong deviations from the expected longitudinal line shape, suggesting that the third order intermodulation is not occurring uniformly in a manner that is simply dictated by the rf current distribution. This is evidence that local information on the sources of nonlinearity are preserved in the IMD PR images [15].

Figure 4 shows a close-up of a 100 μm x 60 μm area along one edge of the CPW center strip near the rf current maximum in the standing wave. Fig. 4(a) shows the ordinary (rf) PR (proportional to $J^2(x,y)$) along the edge and illustrates the edge-current buildup often cited as a potential source of nonlinearity in SC devices [1,2,5,8-10,12,13]. This rf PR was measured by the crystal diode at $f_m = 100$ kHz to provide images with $S = 4.2$ μm. Fig. 4(b) shows IMD PR that was acquired by the spectrum analyzer at $f_m = 6.49$ kHz with $S \sim 15$ μm. (A lower modulation frequency was required to recover the small IMD signal.) The nonlinear response is clearly spread out more in the transverse direction, and is not entirely confined to the edges like the ordinary PR.

To clarify this difference, Fig. 5 compares transverse line cuts through the images shown in Fig. 4. The rf PR peaks just inside the edge, about $S/2 \sim 2$ μm from the edge. The IMD PR peaks 5 μm further inside the film and decays back to zero much more slowly. The spreading of the IMD PR is due (at least in part) to the large thermal healing length $l_T = 15.7$ μm

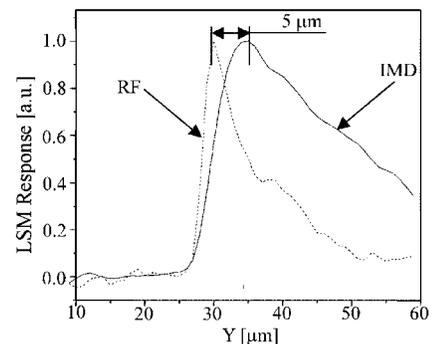

Fig. 5. Normalized Y-profiles of ordinary RF photoresponse (dotted line) and IMD photo response distribution (solid line) averaged through the images shown in Fig.4 (a) and 4(b) respectively. Edge of the resonator is at Y = 28 μm.

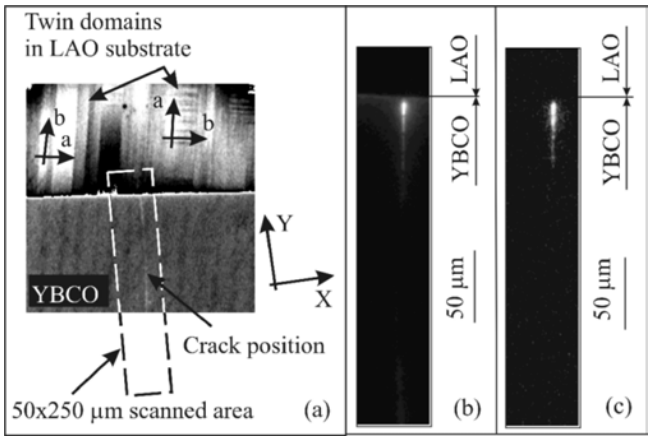

Fig. 6. LSM reflectivity map (a) showing the twinned structure of the substrate and film that can lead to crack appearance. The dashed line shows the area chosen for imaging the (b) RF photoresponse and the (c) IMD photoresponse in a 250 μm long patch centered on the crack.

at the modulation frequency $f_m$ = 6.49 kHz.

Nontrivial behavior of the IMD PR was detected in the vicinity of a crack in the YBCO film that is formed by an area of sharp twin block misorientation, as seen in Fig. 6(a). This behavior is accompanied by an anomalous rf photoresponse peak inside the crack one order of magnitude higher than the ordinary PR from the nearby edge of the film. Both the ordinary rf and IMD PR show a non-bolometric component. Line cuts through these distributions show clear features on length scales smaller than the thermal healing length (Fig.7).

The multiple peaks seen in the IMD PR line cut (Fig. 7) may be due to regions of very large change in local $J_{IMD}$ induced by a direct (nonbolometric) laser depairing mechanism. For instance, these may be pinning sites for rf vortices moving along the crack.

## IV. CONCLUSION

This paper presents the first preliminary IMD photoresponse images on a superconducting microwave device, demonstrating that local nonlinearity imaging is possible with a microwave LSM. Unique contrast is generated by the IMD imaging method. This information is not present in the linear-response bolometric images.

## ACKNOWLEDGMENT

The authors thank K. S. Harshavarden of Neocera, Inc. for making and patterning the YBCO CPW resonators.

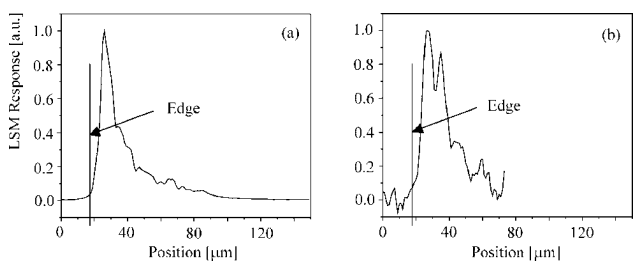

Fig. 7. Line cuts of LSM PR profiles of (a) RF photoresponse and (b) IMD photoresponse distribution along the crack in the YBCO film.

## REFERENCES

[1] Z. Y. Shen, *High-Temperature Superconducting Microwave Circuits* (Artech House, Boston, 1994).
[2] M. Hein, *High-temperature-superconductor thin films at microwave frequencies*, Berlin: Springer, 1999.
[3] G. C. Liang, D. Zhang, C.-F. Shih, M. E. Johansson, R. S. Withers, A. C. Anderson, and D. E. Oates, "High-power high-temperature superconducting microstrip filters for cellular base-station applications," *IEEE Trans. Appl. Supercond.*, vol. 5, pp. 2652–2655, June 1995.
[4] H. Hoshizaki, N. Sakakibara, and Y. Ueno, "Origin of nonlinear surface impedance and intermodulation distortion in $YBa_2Cu_3O_{7-x}$ microstrip resonator," *J. Appl. Phys.* 86, 5788-5793 (1999).
[5] M. A. Golosovsky, H. J. Snortland, and M. R. Beasley, "Nonlinear microwave properties of superconducting Nb microstrip resonators," *Phys. Rev. B* 51, 6462-6469, 1995.
[6] C. Wilker, Z.-Y. Shen, P. Pang, W. L. Holstein, and D. W. Face, "Nonlinear effects in high temperature superconductors: 3rd order intercept from harmonic generation," *IEEE Trans. Appl. Supercond.*, vol. 5, pp. 1665-1670, June 1995.
[7] R. Monaco, A. Andreone and F. Palomba "Intermodulation measurements in Nb superconducting microstrip resonators," *J. Appl. Phys.* 88, pp. 2898-2905, 2000.
[8] T. Dahm and D. J. Scalapino, "Theory of microwave intermodulation in a high-$T_c$ superconducting microstrip resonator," *Appl. Phys. Lett.* 69, pp. 4248-4250, 1996.
[9] J. McDonald, J. R. Clem, D. E. Oates, "Critical-state model for intermodulation distortion in a superconducting microwave resonator," *J. Appl. Phys.* 83, pp. 5307-5312, 1998.
[10] O.G. Vendik, I.B. Vendik, T.B. Samoilova, "Nonlinearity of Superconducting Transmission Line and Microstrip Resonator", *IEEE Trans. Microwave Theory Tech.*, Vol.45, No.2, pp. 173-178, 1997.
[11] J. C. Booth, J. A. Beall, R. H. Ono, F. J. B. Stork, D. A. Rudman, and L. R. Vale, "Third order harmonic generation in high temperature superconducting coplanar waveguides at microwave frequencies," *Appl. Supercond.*, vol. 5, pp. 379–384, August 1998.
[12] J. C. Culbertson, H. S. Newman, and C. Wilker, "Optical probe of microwave current distributions in high temperature superconducting transmission lines," *J. Appl. Phys.* 84, pp. 2768-2787, 1998.
[13] T. Kaiser, M. A. Hein, G. Müller, and M. Perpeet, "Spatially resolved microwave field distribution in YBaCuO disk resonators visualized by laser scanning," *Appl. Phys. Lett.* 73, pp. 3447-3449, 1998.
[14] D. P. Almond, P. Nokrach, E. W. R. Stokes A., Porch, S. A. L. Foulds, F. Wellhofer, J. R. Powell, and J. S. Abell, "Modulated optical reflectance characterization of high temperature superconducting thin film microwave devices," *J. Appl. Phys.* 87, pp. 8628-8635, 2000.
[15] Wensheng Hu, A. S. Thanawalla, B. J. Feenstra, F. C. Wellstood, and Steven M. Anlage, "Imaging of microwave intermodulation fields in a superconducting microstrip resonator", *Appl. Phys. Lett.* 75, pp. 2824-2826, 1999.
[16] P. Lahl, R. Wordenweber, and M. Hein, "Correlation of power handling capability and intermodulation distortion in $YBa_2Cu_3O_{7-\delta}$ thin films," *Appl. Phys. Lett.* 79, pp. 512-514, 2001.
[17] M. Tsindlekht, M. Golosovsky, H. Chayet, D. Davidov, and S. Chocron, "Frequency modulation of the superconducting parallel-plate resonator by laser irradiation", *Appl. Phys. Lett.* 65, pp. 2875-2877, 1994.
[18] D. Quenter, S. Stehle, T. Doderer, C. A. Krulle, R. P. Huebener, F. Muller, J. Niemeyer, R. Popel, T. Weimann, R. Ruby, and A. T. Barfknecht, "Spatially resolved studies of the microwave properties of superconducting devices," *Appl. Phys. Lett.* 63, pp. 2135-2137, 1993.
[19] A. P. Zhuravel, A. V. Ustinov, K. S. Harshavarden, and Steven M. Anlage, "Effect of $LaAlO_3$ Surface Topography on rf Current Distribution in Superconducting Microwave Devices," *Applied Physics Letters*, in press, 16 Dec. 2002.
[20] J. Hartmann, P. Voigt, and M. Reichling, "Measuring local thermal conductivity in polycrystalline diamond with a high resolution photothermal microscope," *J. Appl. Phys.* 81, pp. 2966-2972, 1997.
[21] Rudolf Gross and Dieter Koelle, "Low temperature electron microscopy of superconducting films and Josephson junctions", *Rep. Prog. Phys.*, 57, pp. 651-741, 1994.